\documentclass[10pt]{article}
\usepackage{amssymb}

\begin{document}

\title{Renormalized new solutions for the massless Thirring model}
\author{R. Casana\thanks{casana@ift.unesp.br} \\[0.2cm]
{\it \small Instituto de F\'{\i}sica Te\'orica, Universidade Estadual Paulista}\\
{\it \small Rua Pamplona 145, CEP 01405-900, S\~ao Paulo, SP,
Brazil.}}

\date{} \maketitle

\begin{abstract}
We present a non--perturbative study of the (1+1)--dimensional massless
Thirring model by using path integral methods. The regularization
ambiguities -coming from the computation of the fermionic determinant-
allow to find new solution types for the model. At quantum level the
Ward identity for the 1PI 2-point function for the fermionic current
separates such solutions in two phases or sectors, the first one has a
local gauge symmetry that is implemented at quantum level and the other
one without this symmetry. The symmetric phase is a new solution which
is unrelated to the previous studies of the model and, in the
non--symmetric phase there are solutions that for some values of the
ambiguity parameter are related to well-known solutions of the model.
We construct the Schwinger--Dyson equations and the Ward identities. We
make a detailed analysis of their UV divergence structure and, after,
we perform a non--perturbative regularization and renormalization of
the model.
\end{abstract}


{\small PACs: 03.70.+k, 11.15.-q, 11.30.-j}

%%%%%%%%%%%%%%%%%%%%%%%%%%%%%%%%%%%%%%%%%%%%%%%%%%%%%%%%%%%%%%%%%%%%%

\section{Introduction}

%%%%%%%%%%%%%%%%%%%%%%%%%%%%%%%%%%%%%%%%%%%%%%%%%%%%%%%%%%%%%%%%%%%%%

The massless Thirring model is a well-known exactly solvable quantum field
theory in (1+1) dimensions, as it has been shown, for the first time, by
Thirring in his seminal paper \cite{thirring1}. Thirring constructed the
eigenstates of the Hamiltonian and calculated some observable quantities and
found them to be finite after renormalization. From that date, many studies
and extensive investigation of the model have been done. Hagen \cite{hagen1}
introduced an external field and found the general solution of the Thirring
model. Of course, the exact solution of the massless Thirring model means
the possibility of an explicit evaluation of any fermionic correlation
function, and it was carried out, for the first time, by Hagen \cite{hagen1}%
, and afterwards by Klaiber \cite{klaiber1}. In the Klaiber's analysis of
the Thirring model, he found the operator solutions which are expressed in
terms of a free massless Dirac field. On the other hand, Nakanishi \cite%
{nakanishi} expressed the solution in terms of the free massless bosonic
field and asserted that all Heisenberg operators should be expressed in
terms of asymptotic fields from the standpoint of the general principles of
quantum field theory.

Recently a study has been made on the massless Thirring model has
been made whose purpose was to investigate the equivalence between
the massive Thirring model and the sine-Gordon model \cite{faber1}.
On the other hand an extensive and detailed analysis of the model in
its chiral symmetric phase was performed in \cite{faber2} and its
operator bosonic representation in \cite{faber3}.

In previous works, the ambiguities have been observed in the fermionic
current definition, as it was noted by Johnson \cite{Johnsonx}, Hagen \cite%
{hagen1} and Klaiber \cite{klaiber1}, and it affects the coupling
constant, therefore, in the massless Thirring model, the coupling
constant is determined only when we define the current
regularization. It was proved that the coupling constants are
related to each other depending on the regularization. For example
the coupling constant of the Schwinger definition \cite{Scwingerx}
$g_{_S}$ is given by
\begin{eqnarray}
g_{_S}=\frac{g_{_J}}{1-\frac{g_{_J}}{2\pi}}\;,  \nonumber
\end{eqnarray}
where $g_{_J}$ is the coupling constant of the Johnson definition \cite%
{Johnsonx}. Another study of ambiguity of the current regularization was
performed in \cite{takahashi1}. These current ambiguities also appear in the
massive Thirring model which still remains to be understood \cite{fujita1}.

The aim of the present work is to show renormalized new solutions
for the model obtained by using non--perturbative techniques via the
functional integral method. The key to get such solutions are the
regularization ambiguities that appear in the calculation of the
fermionic determinant. Once the fermionic determinant is
well--defined we are capable to solve the model without having to
define the fermionic current \cite{Jackiwtop,banerjee1}.

Then, the regularization ambiguities allow us to show that, at
quantum level, there are two ways to get the full quantization of
the massless Thirring model. The first one has solutions that are
related to the various studies of the model just performed over the
last years in which they do not consider the possibility of
implementing a local gauge symmetry at the quantum level by
exploring the ambiguities that appear in the calculation of the
fermionic determinant or in the definition of the fermionic current.
And the second one is a solution, as it will be shown, which results
when we impose a local gauge symmetry at the quantum level. This
imposition leads to an equation which fixes the ambiguity parameter
as a simple function of the coupling constant. The unexpected result
comes from the requirement of a local gauge symmetry which is not
present at the classical level.

The paper is displayed in the following way. In section 2, we
present the model and shown the consequences of the regularization
ambiguities in the solutions for the model. The section 3 and 4
shows a detailed analysis of the full quantization of the massless
Thirring model in both phases mentioned in the previous paragraph.
We do an extensive study about the calculus of the relevant Green's
functions, the construction of the Schwinger--Dyson equations and
Ward identities. Moreover, we make a careful analysis of the
divergence structures and their regularization procedures, and
finally we establish the renormalization and discuss their
consequences. In section 5, we give our conclusion and perspectives.

We use the following conventions:
\[
g^{00}=1=-g^{11}\,,\; \epsilon^{01}=1=-\epsilon_{01}\;,
\]
\[
\{\gamma^\mu, \gamma^\nu\} =2\,g^{\mu\nu}\;,\quad \{\gamma_\mu,\gamma_5\}=0
, \quad\gamma_5=\gamma^0 \gamma^1\;, \quad P_\pm=\frac{1}{2}(1\pm\gamma_5).
\]
\[
\tilde V^\mu= \epsilon^{\mu\nu}V_\nu.
\]

%%%%%%%%%%%%%%%%%%%%%%%%%%%%%%%%%%%%%%%%%%%%%%%%%%%%%%%%%%%%%%%%%%%%%

\section{The massless Thirring model}

%%%%%%%%%%%%%%%%%%%%%%%%%%%%%%%%%%%%%%%%%%%%%%%%%%%%%%%%%%%%%%%%%%%%%

The massless Thirring model \cite{thirring1} is a theory of a self--coupled
Dirac field defined by the following Lagrangian density
\begin{equation}
\mathcal{L}[\bar{\psi},\psi ]=\bar{\psi}{i}\partial \!\!\!\slash\psi -\frac{g%
}{2}\;(\bar{\psi}\gamma ^{\mu }\psi )^{2}\;,  \label{th-eqx1}
\end{equation}%
where $g\neq 0$. The Lagrangian is invariant under the global chiral group $%
U_{_{V}}(1)\times U_{_{A}}(1)$ at classical level,
\begin{equation}
\psi \longrightarrow e^{i\alpha }\psi \;,\quad \psi \longrightarrow
e^{i\beta \gamma _{_{5}}}\psi \,.  \label{th-eqx2}
\end{equation}

We quantize this fermionic model following the functional integral method,
thus we write its generating functional as
\begin{equation}
Z[\eta ,\overline{\eta }]=\int \!\!\mathcal{D}\psi \mathcal{D}\bar{\psi}%
\;\exp \left( i\int \!\!dx\;\bar{\psi}{i}\partial \!\!\!\slash\psi -\frac{g}{%
2}\;(\bar{\psi}\gamma ^{\mu }\psi )^{2}+\bar{\eta}\psi +\bar{\psi}\eta
\right) .  \label{th-eqx3}
\end{equation}%
The first step in the quantization procedure is to linearize the quadratic
term in the fermionic current introducing an auxiliary vector field $A_{\mu
} $, such that the generating functional is expressed as
\begin{equation}
Z[\eta ,\bar{\eta}]=\!\int \!\!\mathcal{D}A_{\mu }\mathcal{D}\psi \mathcal{D}%
\bar{\psi}\;\exp \left( i\!\int \!\!dx\;\bar{\psi}(i\partial \!\!\!\slash%
+A\!\!\!\slash)\psi +\frac{1}{2g}A_{\mu }A^{\mu }+\bar{\eta}\psi +\bar{\psi}%
\eta \right) \;,  \label{th-eqx4}
\end{equation}%
where the presence of the massive term for the vector field precludes the
existence (at classical level) of the following $U(1)$ local gauge symmetry,
\begin{equation}
\psi \longrightarrow e^{i\alpha }\psi \;,\quad \bar{\psi}\longrightarrow
\bar{\psi}e^{-i\alpha }\;,\quad A_{\mu }\longrightarrow A_{\mu }+\partial
_{\mu }\alpha \;.  \label{th-eqx5}
\end{equation}

The fact that at classical level the model does not have a local gauge
symmetry allows us to compute the fermionic determinant using a
generalized regularization prescription
\cite{Jackiwtop,banerjee1,Harada,DiasLinhares1,mitra1} to control the
UV divergences. By following the point-splitting regularization
implemented in \cite{DiasLinhares1}, we compute the fermionic
determinant to be
\begin{equation}
\det (i\partial \!\!\!\slash+A\!\!\!\slash)=\exp \left( i\!\int \!\!dx\;%
\frac{1}{2\pi }A_{\mu }\left[ \frac{a+1}{2}g^{\mu \nu }-\frac{\partial ^{\mu
}\partial ^{\nu }}{\square }\right] A_{\nu }\right)  \label{th-eqx6}
\end{equation}%
where the parameter $a$ characterizes the different regularization
procedures used to control the UV divergences. When $a=1$ the fermionic
determinant is invariant under the local gauge transformation
$A_{\mu}\rightarrow A_{\mu}+\partial_{\mu }\alpha $.

In the following, we study the model for all the possible values of the
ambiguity parameter $a\neq 1$.

After the fermionic integration the generating functional (\ref{th-eqx4})
reads as
\begin{equation}
Z[\eta ,\bar{\eta}]=\!\int \!\!\mathcal{D}A_{\mu }\;\exp \left( i\!\int
\!\!dx\;\frac{1}{2\pi }\,A_{\mu }\left[ \left( \frac{\pi }{g}+\frac{a+1}{2}%
\right) \;g^{\mu \nu }-\frac{\partial ^{\mu }\partial ^{\nu }}{\square }%
\right] A_{\nu }\right) \;\exp \left( -i\!\int \!\!dx\,dy\;\bar{\eta}%
(x)G(x,y;A)\eta (y)\right) \,,  \label{th-eqx7}
\end{equation}%
where $G(x,y;A)$ is the Green's function of the Dirac's equation: $\left(
i\partial \!\!\!\slash+A\!\!\!\slash\right) G(x,y;A)=\delta (x-y)$, and it
can be exactly computed,
\begin{equation}
G(x,y;A)=\exp \left( -i\int \!\!dz\;A_{\mu }(z)j_{+}^{\mu }(z,x,y)\right)
P_{+}G_{F}(x-y)+\exp \left( -i\int \!\!dz\;A_{\mu }(z)j_{-}^{\mu
}(z,x,y)\right) P_{-}G_{F}(x-y)\,,  \label{th-eqx9}
\end{equation}%
with $G_{F}(x-y)$ being the Green's function of the massless Dirac's
equation: $\;i\partial \!\!\!\slash G_{F}(x-y)=\delta (x-y)$; and the
contact current $j_{\pm }^{\mu }(z,x,y)$ is
\begin{equation}
j_{\pm }^{\mu }(z,x,y)=(\partial _{z}^{\mu }\mp \tilde{\partial}_{z}^{\mu
})[D_{F}(z-x)-D_{F}(z-y)]\;,  \label{th-eqx10}
\end{equation}%
where $D_{F}(x)$ is the Green's function of the massless Klein-Gordon
equation: $\;\square D_{F}(x-y)=\delta (x-y)\,$.

\vskip0.3cm At quantum level the Ward identity for the 1PI 2-point function
of the fermionic current is given by%
\begin{equation}
q_{\mu }\Gamma ^{\mu \nu }(q)=-\left( g+\displaystyle{\frac{2\pi }{a-1}}%
\right) q^{\nu }\;.  \label{th-eqxx10}
\end{equation}
then, we will have a local $U\left( 1\right)$ gauge symmetry if and
only if
\begin{equation}
a=1-\displaystyle\frac{2\pi }{g}  \label{th-eqx12}
\end{equation}
such symmetry is reflected in the effective action (\ref{th-eqx7})
which is invariant under the local $U(1)$ gauge transformation
\begin{equation}
A_{\mu }\rightarrow A_{\mu }+\partial _{\mu }\alpha \,.  \label{th-eqx11}
\end{equation}%
The equation (\ref{th-eqx12}) is important since the ambiguity
becomes a simple function of the fundamental coupling of the model.
The ambiguity that appears in the regularization process of the
fermionic determinant or in the definition of the fermionic current
is fixed by the requirement of a new local gauge symmetry at quantum
level. Therefore, the model at quantum level displays two phases:
one of them does not have the local $U(1)$ gauge symmetry and the
other one has. In the following we analyze both phases separately.

%%%%%%%%%%%%%%%%%%%%%%%%%%%%%%%%%%%%%%%%%%%%%%%%%%%%%%%%%%%%%%%%%%%%%

\section{The non--gauge invariant case: $a \neq 1-\frac{2\protect\pi}{g}$}

%%%%%%%%%%%%%%%%%%%%%%%%%%%%%%%%%%%%%%%%%%%%%%%%%%%%%%%%%%%%%%%%%%%%%

We begin the quantization process from the generating functional defined in (%
\ref{th-eqx4}) and we include the external source for the fermionic current,
thus, we have
\begin{equation}
Z[\eta ,\bar{\eta},C_{\mu }]=\int \!\!\mathcal{D}A_{\mu }\mathcal{D}\psi
\mathcal{D}\bar{\psi}\;\exp \left( i\int \!\!dx\;\bar{\psi}(i\partial \!\!\!%
\slash+A\!\!\!\slash+C\!\!\!\!\slash)\psi +\frac{1}{2g}A_{\mu }A^{\mu }+\bar{%
\eta}\psi +\bar{\psi}\eta \right) .  \label{th-eqx13}
\end{equation}%
In this situation we do not need a gauge fixing term for the vector field
because its effective action (\ref{th-eqx7}) is not gauge invariant under (%
\ref{th-eqx11}). Thus, considering the fermionic determinant in (\ref%
{th-eqx6}) for $a\neq 1-\frac{2\pi }{g}$ we can compute all the
correlation functions and we can also construct the Schwinger-Dyson
equations (SDE) and the Ward identities.

%%%%%%%%%%%%%%%%%%%%%%%%%%%%%%%%%%%%%%%%%%%%%%%%%%%%%%%%%%%%%%%%%%%%%

\subsection{The Green's functions}

%%%%%%%%%%%%%%%%%%%%%%%%%%%%%%%%%%%%%%%%%%%%%%%%%%%%%%%%%%%%%%%%%%%%%

\subsubsection{The fermion propagator $G(x-y)=\langle 0|T\psi(x)
\bar{\psi}(y)|0\rangle $}

The fermion propagator reads as
\begin{equation}
G(x-y)=i\,\exp \left( -\frac{i\,\pi }{b(b-1)}\int \!\!\frac{dk}{(2\pi )^{2}}%
\;\frac{1-e^{-ik\cdotp(x-y)}}{k^{2}}\right) G_{F}(x-y)\;,  \label{th-eqx14}
\end{equation}%
w\bigskip here
\begin{equation}
b=\frac{\pi }{g}+\frac{a+1}{2}  \label{th-eqx14-a}
\end{equation}

The fermion propagator has ultraviolet divergences and it will be shown that
after its regularization and renormalization procedures guarantee the full
finiteness of the model.

\subsubsection{The current propagator $J_{\mu\nu}(x-y)=\langle 0|Tj_{\mu}
(x)j_{\nu}(y)|0\rangle $}

In momentum space\ the propagator of the fermionic current ${j^{\mu }}=\bar{%
\psi}\gamma ^{\mu }\psi $ is calculated to be
\begin{equation}
i\,J_{\mu \nu }(q)=\frac{T_{\mu \nu }}{g+\displaystyle{\frac{2\pi }{a+1}}}+%
\frac{L_{\mu \nu }}{g+\displaystyle{\frac{2\pi }{a-1}}}  \label{th-eqx15}
\end{equation}%
and it is free of UV divergences and, the tensors $T_{\mu \nu }$ and $%
L_{\mu \nu }$ are the transversal and longitudinal projectors defined as
\begin{equation}
T_{\mu \nu }=g_{\mu \nu }-\frac{q_{\mu }q_{\nu }}{q^{2}}\quad ,\quad %
\mbox{and}\quad L_{\mu \nu }=\frac{q_{\mu }q_{\nu }}{q^{2}}\;.
\label{th-eqx16}
\end{equation}

\subsubsection{The vertex function $G^{\protect\mu }(x,y;z)=\langle
0|T\psi (x)\bar{\psi}(y)j^{\mu}(z)|0\rangle $}

It is calculated to be
\begin{equation}
{G}^{\mu }(x,y;z)=i\int \!\!\frac{dk}{(2\pi )^{2}}\left( -\frac{\pi }{g(b-1)}%
\frac{k^{\mu }}{k^{2}}+\frac{\pi }{gb}\frac{\tilde{k}^{\mu }}{k^{2}}\,\gamma
_{5}\right) \left( e^{-ik\cdotp(z-x)}-e^{-ik\cdotp(z-y)}\right) G(x-y).
\label{th-eqx17}
\end{equation}%
In momentum space it reads as
\begin{equation}
G^{\mu }(p,-p-q;q)=i\left( -\frac{\pi }{g(b-1)}\frac{q^{\mu }}{q^{2}}+\frac{%
\pi }{gb}\frac{\tilde{q}^{\mu }}{q^{2}}\,\gamma _{5}\right) \left[
\,G(p+q)-G(p)\frac{{}}{{}}\!\right] .  \label{th-eqx18}
\end{equation}%
where $G(p)$ is the Fourier transforms of the fermionic propagator (\ref%
{th-eqx14}). Thus, we can note that the UV divergences of this Green`s
function come from the fermionic propagator.

It can be shown that $2n$-point Green's functions for the pure fermionic
current are UV finite, and all the $2n+m$-point Green's functions involved $%
2n$ fermions and $m$ current insertions have logarithmic UV
divergences.

%%%%%%%%%%%%%%%%%%%%%%%%%%%%%%%%%%%%%%%%%%%%%%%%%%%%%%%%%%%%%%%%%%%%%

\subsubsection{The 1PI functions}

%%%%%%%%%%%%%%%%%%%%%%%%%%%%%%%%%%%%%%%%%%%%%%%%%%%%%%%%%%%%%%%%%%%%%
We write the 1PI in momentum space. Thus, the 1PI 2-point fermion
function $\Gamma (p)$ which is the inverse of the fermionic propagator
is defined as
\begin{equation}
\Gamma (p)=iG^{-1}(p)\;.  \label{th-eqx19}
\end{equation}%

The 1PI function 2-point function of the fermion current $\Gamma ^{\mu
\nu }(q)$ is given by
\begin{eqnarray}
\Gamma _{\mu \nu }(q) =i\,J_{\mu \nu }^{-1}(q)= -\left(
g+\displaystyle{\frac{2\pi }{a+1}}\right) T_{\mu \nu }-\left(
g+\displaystyle{\frac{2\pi }{a-1}}\right) L_{\mu \nu }
 \label{th-eqx20-a}
\end{eqnarray}

The 1PI 3-point function $\Gamma ^{\mu }(p,k;q)$ related to the vertex
function $G^{\mu }(p,k;q)$ is
\begin{equation}
\Gamma ^{\mu }(p,-p-q;q)=\left( \frac{2\pi }{a-1}\,\frac{q^{\mu }}{q^{2}}+%
\frac{2\pi }{a+1}\,\frac{\tilde{q}^{\mu }}{q^{2}}\;\gamma _{5}\right) \left[
\Gamma (p+q)\frac{{}}{{}}\!\!\,-\Gamma (p)\right]  \label{th-eqq21}
\end{equation}

%%%%%%%%%%%%%%%%%%%%%%%%%%%%%%%%%%%%%%%%%%%%%%%%%%%%%%%%%%%%%%%%%%%%%

\subsection{Schwinger--Dyson equations}

%%%%%%%%%%%%%%%%%%%%%%%%%%%%%%%%%%%%%%%%%%%%%%%%%%%%%%%%%%%%%%%%%%%%%

We start from the generating functional (\ref{th-eqx13}) that after
integration of the vector field can reads as
\begin{equation}
Z[\eta ,\bar{\eta},C_{\mu }]=\!\int \!\!\mathcal{D}\psi \mathcal{D}\bar{\psi}%
\;\exp \left( i\!\int \!\!dx\;\bar{\psi}(i\partial \!\!\!\slash+C\!\!\!\slash%
)\psi -\frac{g}{2}(\bar{\psi}\gamma ^{\mu }\psi )^{2}+\bar{\eta}\psi +\bar{%
\psi}\eta \right) \;,  \label{th-eqq22}
\end{equation}

Making the functional variation of the generating functional with
respect to the fermionic field $\bar{\psi}$ we get the quantum equation
of motion
\begin{equation}
\left( \partial _{x}\!\!\!\!\!\!\slash\;\;\frac{\delta }{\delta \bar{\eta}(x)%
}+g\gamma ^{\mu }\frac{\delta }{\delta \bar{\eta}(x)}\frac{\delta }{\delta
C^{\mu }(x)}-iC_{\mu }(x)\gamma ^{\mu }\frac{\delta }{\delta \bar{\eta}(x)}%
+\eta (x)\right) Z[\eta ,\bar{\eta},C_{\mu }]=0  \label{th-eqx22}
\end{equation}%
where we have used the functional representations for the fields
\begin{equation}
\psi =-i\frac{\delta }{\delta \bar{\eta}}\quad ,\quad \bar{\psi}=i\frac{%
\delta }{\delta \eta }\quad ,\quad j^{\mu }=-i\frac{\delta }{\delta
C_{\mu }}. \label{th-eqq23}
\end{equation}
We perform a functional derivative with respect to $\eta (y)$ in the
functional equation (\ref{th-eqx22}) and setting the external sources
to zero we obtain the SDE for the fermion propagator
\begin{equation}
i\partial _{x}\!\!\!\!\!\!\slash\;\;G(x-y)=i\delta (x-y)+g\gamma ^{\mu
}G_{\mu }(x,y;x)  \label{th-eq23}
\end{equation}%
which is satisfied by the Green's functions computed in (\ref{th-eqx14}) and
(\ref{th-eqx17}) from the generating functional (\ref{th-eqx13}) when we
consider the fermionic determinant (\ref{th-eqx6}) for $a\neq 1-\frac{2\pi }{%
g}$.

As we mentioned above, the fermion propagator has UV divergences that
are shown explicitly when we write the Schwinger-Dyson equation in
momentum space
\begin{equation}
G(p)=\frac{i}{p\!\!\!\slash}\;-\;i\int \!\!\frac{dk}{(2\pi )^{2}}\;f(k)\;%
\frac{1}{p\!\!\!\slash}\;k\!\!\!\slash\;G(p-k)\;.  \label{th14}
\end{equation}%
where the function $f(k)$ reads for
\begin{equation}
f(k)=-\;\frac{\pi }{b(b-1)\;k^{2}}\;  \label{th13}
\end{equation}%
Iterating the SDE we generate an expansion in $\hbar $ or equivalently in
powers of $f(k)$
\begin{equation}
G(p)=\frac{i}{p\!\!\!\slash}\;+\int \!\!\frac{dk}{(2\pi )^{2}}\;f(k)\;\frac{1%
}{p\!\!\!\slash}\;k\!\!\!\slash\;\frac{1}{p\!\!\!\slash-k\!\!\!\slash}%
\;-\;i\int \!\!\frac{dk}{(2\pi )^{2}}\frac{dq}{(2\pi )^{2}}\;f(k)f(q)\;\frac{%
1}{p\!\!\!\slash}\;k\!\!\!\slash\;\frac{1}{p\!\!\!\slash-k\!\!\!\slash}%
\;q\!\!\!\slash\;\frac{1}{p\!\!\!\slash-k\!\!\!\slash-q\!\!\!\slash}%
\;+\;\ldots  \label{th14x}
\end{equation}%
which is a semi--perturbative expansion just as it was shown in \cite{trab1}
for the anomalous and chiral Schwinger model cases. A power counting
analysis of the series clearly shows the existence of a logarithmic UV
divergence in each order or loop, such property of the series guarantees
that the model is a fully renormalizable field theory.

%%%%%%%%%%%%%%%%%%%%%%%%%%%%%%%%%%%%%%%%%%%%%%%%%%%%%%%%%%%%%%%%%%%%%

\subsection{The Ward identities}

%%%%%%%%%%%%%%%%%%%%%%%%%%%%%%%%%%%%%%%%%%%%%%%%%%%%%%%%%%%%%%%%%%%%%

We start from the generating functional (\ref{th-eqx13}) and we make
the following fermionic local phase transformation
\begin{equation}
\psi \rightarrow e^{i\lambda (x)}\psi \quad ,\quad \bar{\psi}
\rightarrow \bar{\psi}e^{-i\lambda (x)}\,.  \label{th-eqx25}
\end{equation}
In our framework the fermionic measure is not invariant under such
transformation changing as
\begin{equation}
\mathcal{D}\psi \mathcal{D}\bar{\psi}\rightarrow J[A,\lambda]
\,\mathcal{D}\psi \mathcal{D}\bar{\psi}
\end{equation}
where $J[A,\lambda]$ is the Jacobian given by
\begin{equation}
J[A,\lambda]\;=\;\frac{\det \left( {i}\partial
+A\!\!\!\slash+C\!\!\!\!\slash\right)
}{\det \left( {i}\partial +A\!\!\!\slash+C\!\!\!\!\slash-\partial \!\!\!%
\slash\lambda \right) }\;=\;\exp \left( i\int \!\!dx\;\frac{a-1}{2\pi }%
\;(A_{\mu }+C_{\mu })\partial ^{\mu }\lambda -\frac{a-1}{4\pi }\;(\partial
_{\mu }\lambda )^{2}\right) ,
\end{equation}
then, the generating functional in (\ref{th-eqx13}) reads as
\begin{equation}
Z[\eta ,\bar{\eta},C_{\mu }]\;=\int \!\!\mathcal{D}A_{\mu }\mathcal{D}\psi
\mathcal{D}\bar{\psi}\,J[A]\,\exp \left( i\int \!\!dx\;\bar{\psi}\left( {i}%
\partial +A\!\!\!\slash+C\!\!\!\!\slash-\partial \!\!\!\slash\lambda \right)
\psi +\frac{1}{2g}\,A_{\mu }A^{\mu }+\bar{\eta}e^{i\lambda (x)}\psi +\bar{%
\psi}e^{-i\lambda (x)}\eta \right) .
\end{equation}

Next, we do the following transformation in the vector field $A_{\mu
}\rightarrow A_{\mu }+\partial _{\mu }\lambda $ and after
integrating it we obtain
\begin{eqnarray}
Z[\eta ,\bar{\eta},C_{\mu }] &=&\!\!\int \!\!\mathcal{D}\psi \mathcal{D}\bar{%
\psi}\;\exp \left( i\int \!\!dx\;\bar{\psi}{i}\partial \!\!\!\slash\psi -%
\frac{g}{2}\;(\bar{\psi}\gamma ^{\mu }\psi )^{2}+\bar{\eta}e^{i\lambda
(x)}\psi +\bar{\psi}e^{-i\lambda (x)}\eta +C_{\mu }j^{\mu }\right)
\label{th-eqx28} \\
&&\exp \left( i\int \!\!dx\left[ \frac{a-1}{2\pi }\,C_{\mu }-g\,\frac{b-1}{%
\pi }\,j_{\mu }\right] \partial ^{\mu }\lambda +\frac{1}{2}\left[ \frac{b-1}{%
\pi }-g\,\frac{(b-1)^{2}}{\pi ^{2}}\right] (\partial _{\mu }\lambda
)^{2}\right)  \nonumber.
\end{eqnarray}

By considering $\lambda (x)$ as an infinitesimal function and expanding
to first order in $\lambda$, we get the functional equation satisfied
by the generating functional (\ref{th-eqx13})
\begin{equation}
\left( \bar{\eta}\,\frac{\delta }{\delta \bar{\eta}}-\eta \,\frac{\delta }{%
\delta \eta }-\frac{a-1}{2\pi }\,\partial ^{\mu }{C}_{\mu }-ig\,\frac{b-1}{%
\pi }\,\partial ^{\mu }\frac{\delta }{\delta {C}^{\mu }}\right) Z[\eta ,\bar{%
\eta},C_{\mu }]=0
\end{equation}%
where we have used the functional representation (\ref{th-eqq23}) for
the fields.

The generating functional of the connected Green's functions $W[\eta ,\bar{%
\eta},C_{\mu }]$ is defined as being $W=-i\ln Z$ and the generating
functional of the 1PI functions $\Gamma \lbrack \bar{\psi},\psi ,j^{\mu
}]$ , both are related via the following Legendre transformation
\begin{equation}
\Gamma \lbrack \bar{\psi},\psi ,j^{\mu }]=W[\eta ,\bar{\eta},C_{\mu }]-\int
\!\!dx\left( \bar{\eta}\psi +\bar{\psi}\eta +C_{\mu }j^{\mu }\right) ,
\label{th-eqx30}
\end{equation}%
and using the functional relations
\begin{equation}
\psi =\frac{\delta W}{\delta \bar{\eta}}\quad ,\quad \bar{\psi}=-\frac{%
\delta W}{\delta \eta }\quad ,\quad j_{\mu }=\frac{\delta W}{\delta C^{\mu }}%
\quad ,\quad \bar{\eta}=\frac{\delta \Gamma }{\delta \psi }\quad ,\quad \eta
=-\frac{\delta \Gamma }{\delta \bar{\psi}}\quad ,\quad C_{\mu }=\frac{\delta
\Gamma }{\delta j^{\mu }}\,,  \label{th-eqx31}
\end{equation}%
we obtain the functional equation satisfied by the generating functional $%
\Gamma \lbrack \bar{\psi},\psi ,j^{\mu }]$ from which we get all Ward
identities
\begin{equation}
i\,\frac{\delta \Gamma }{\partial \psi (x)}\,\psi
(x)-i\,\frac{\delta \Gamma }{\partial
\bar{\psi}(x)}\,\bar{\psi}(x)+\frac{a-1}{2\pi }\,\partial _{x}^{\mu
}\,\frac{\delta \Gamma }{\partial {j}^{\mu }(x)}+g\,\frac{b-1}{\pi
}\;\partial _{x}^{\mu }\,{j}_{\mu }(x)=0  \label{th-eqx32}.
\end{equation}

\vskip 0.5cm

The first Ward identity is for the 1PI 2-point function $\Gamma ^{\mu \nu
}(q)$ of the fermionic current which reads as
\begin{equation}
q_{\mu }\Gamma ^{\mu \nu }(q)=-\left( g+\displaystyle{\frac{2\pi }{a-1}}%
\right) q^{\nu }\;.  \label{th-eqx33-a}
\end{equation}%
as it can be shown explicitly from the equation (\ref{th-eqx20-a}). We
can note that the current propagator is transverse for the critical
value $a=1-\displaystyle{\frac{2\pi }{g}}\;\;$ and, as we mentioned in
Section 2, it implies in the existence of a local $U(1)$ gauge symmetry
for this value of the ambiguity parameter.

The second Ward identity which relates the 1PI 3-point function $\Gamma
^{\mu }(p,-p-q;q)$ and the 1PI 2-point fermion function $\Gamma (p)$ \ is
given by
\begin{equation}
q_{\mu }\Gamma ^{\mu }(p;q)=\frac{2\pi }{a-1}\left[ \Gamma (p+q)\frac{{}}{{}}%
\!\!\,-\Gamma (p)\right] ,  \label{th-eqx34}
\end{equation}%
it can also be shown easily from the equation (\ref{th-eqq21}). This
identity guarantees that all the divergences come from the fermion
propagator. Thus, we can see that the ambiguity parameter does not
renormalize and consequently the coupling constant either.

%%%%%%%%%%%%%%%%%%%%%%%%%%%%%%%%%%%%%%%%%%%%%%%%%%%%%%%%%%%%%%%%%%%%%

\subsection{Regularization and renormalization of the UV divergences \label%
{regbneq1}}

%%%%%%%%%%%%%%%%%%%%%%%%%%%%%%%%%%%%%%%%%%%%%%%%%%%%%%%%%%%%%%%%%%%%%

We will study the UV divergences structure of the pure fermionic
$2n$-point Green`s functions (correlation functions without current
insertions) to identify their origin and later to perform their
regularization. The $(2n+m)$--point Green's functions, where $m$ is
the number of current insertions, have similar divergence structure.
Thus, as we will see, a simple renormalization of fermionic wave
function is sufficient for all Green's functions becomes finite.

Then, in the generating functional (\ref{th-eqx7}) we decompose the
vector field $A_{\mu }$ in their longitudinal and transversal
components
\begin{equation} A_{\mu }\;=\;\partial _{\mu }\rho
-\tilde{\partial}_{\mu }\phi \;, \label{th-eqx40}
\end{equation}%
getting the following expression for the generating functional
\begin{equation}
Z[\eta ,\bar{\eta}]=\int \!\!\mathcal{D}\rho \mathcal{D}\phi \,\exp \!\left(
i\!\int \!\!dx\;\frac{b-1}{2\pi }\,\partial _{\mu }\rho \partial ^{\mu }\rho
-\frac{b}{2\pi }\,\partial _{\mu }\phi \partial ^{\mu }\phi \right) \;\exp
\left( -i\!\int \!dx\,dy\,\bar{\eta}(x)G(x,y;\rho ,\phi )\eta (y)\right) ,
\label{th18}
\end{equation}%
where $G(x,y;\rho ,\phi )$ is given by
\begin{equation}
G(x,y;\rho ,\phi )=\sum_{\epsilon =\pm }\exp \left( i\int \!\!dz~\left[
\!\rho (z)-\epsilon \phi (z)\right] j(z,x,y)\right) P_{\epsilon
}\,G_{F}(x-y)\;,  \label{th-eqx42}
\end{equation}%
and the contact current $j(z,x,y)=\delta (z-x)-\delta (z-y)$. From the
generating functional (\ref{th18}) we can compute all pure fermionic
Green's functions, for example, the fermionic propagator
\begin{eqnarray}
G(x-y) &=&i\int \mathcal{D}\rho ~\exp \left( i\int\!\!dz\; \frac{b-1}{2\pi}
\;\partial _{\mu }\rho \partial ^{\mu }\rho +\rho (z)j(z,x,y)\right) \, \\
&&\sum_{\epsilon =\pm }\int \mathcal{D}\phi ~\exp \left( -i\int
\!\!dz\; \frac{b}{2\pi }\;\partial _{\mu }\phi \partial ^{\mu }\phi
+\epsilon \phi (z)j(z,x,y)\right) \,P_{\epsilon }\,G_{F}(x-y)\;.
\nonumber
\end{eqnarray}
We can note that the functional integration on the fields $\rho $
and $\phi $ are factorized. After performing the functional
integration we can see that the ultraviolet divergences appear due
to the bad behavior at high energy of the propagator of the
transversal and longitudinal components of the vector field. The
same situation happened when we compute all others higher order
$2n$--point and $(2n+m)$--point Green's functions.

As we have already known the origin of the ultraviolet divergences,
our next step will be to construct a non--perturbative scheme of
regularization inside the generating functional. Thus, we begin the
regularization process from the generating functional
(\ref{th-eqx13}) setting $C_{\mu }=0$ which reads
\begin{equation}
Z[\eta ,\bar{\eta}]=\int \!\!\mathcal{D}A_{\mu }\mathcal{D}\psi \mathcal{D}%
\bar{\psi}\;\exp \left( i\int \!\!dx\;\bar{\psi}\left( {i}\partial +A\!\!\!%
\slash\right) \psi +\frac{1}{2g}\;A_{\mu }A^{\mu }+\bar{\eta}\psi +\bar{\psi}%
\eta \right) .  \label{th-eqx43}
\end{equation}%
We decompose the vector field as in (\ref{th-eqx40}) and make the following
fermionic transformation
\begin{equation}
\psi \rightarrow \,e^{i\rho -i\gamma _{5}\phi }\psi \quad ,\quad \;\bar{\psi}%
\rightarrow \bar{\psi}\,e^{-i\rho -i\gamma _{5}\phi }\,.  \label{th22}
\end{equation}%
As it is well--known, the fermionic measure is not invariant under a chiral
transformation \cite{fujikawa1}. On the other hand, we can also show that it
is not invariant under a phase transformation \cite{Harada,trab2} due to the
generalized prescription used to compute the fermionic determinant. Thus,
the fermionic measure under the transformation (\ref{th22}) changes as $%
\mathcal{D}\psi \mathcal{D}\bar{\psi}\rightarrow \,J[\rho ,\phi ]\mathcal{D}%
\psi \mathcal{D}\bar{\psi}$, where $J[\rho ,\phi ]$ is the Jacobian given by
\begin{equation}
J[\rho ,\phi ]=\frac{\det (i\partial \!\!\!\slash+\partial \!\!\!\slash\rho -%
\tilde{\partial}\!\!\!\slash\phi )}{\det (i\partial \!\!\!\slash)}=\exp
\left( i\!\int \!\!dx\;\frac{a-1}{4\pi }\;\partial _{\mu }\rho \partial
^{\mu }\rho -\frac{a+1}{4\pi }\;\partial _{\mu }\phi \partial ^{\mu }\phi
\right) \,.  \label{th22-1}
\end{equation}

Inserting (\ref{th-eqx40}), (\ref{th22}) and (\ref{th22-1}) into the
generating functional (\ref{th-eqx43}), we get
\begin{equation}
Z[\eta ,\bar{\eta}]\,=\!\int \!\!\mathcal{D}\rho \mathcal{D}\phi \,\mathcal{D%
}\psi \mathcal{D}\bar{\psi}\;\exp \left( i\int \!\!dx\;\bar{\psi}i\partial
\!\!\!\slash\psi +\frac{b-1}{2\pi }\;\partial _{\mu }\rho \partial ^{\mu
}\rho -\frac{b}{2\pi }\;\partial _{\mu }\phi \partial ^{\mu }\phi +\bar{\eta}%
e^{i\rho -i\gamma _{5}\phi }\psi +\bar{\psi}e^{-i\rho -i\gamma _{5}\phi
}\eta \right) .  \label{th-eqx47}
\end{equation}%
The \textit{fundamental} observation here, is that we could make everything
finite if we had a better UV behavior for the $\rho $ and $\phi $
propagators. Then, by following the scheme showed in \cite{trab1}, we add to
the generating functional two regulator fields $\alpha $ and $\beta $ with
large masses $\Lambda _{_{\alpha }}$ and $\Lambda _{_{\beta }}$ respectively
\cite{Zinn-Justin} ($\Lambda ^{2}\rightarrow +\infty $). Also, the
interacting terms of $\rho $ and $\phi $ fields are changed as $\rho
\rightarrow \rho +\alpha $ and $\phi \rightarrow \phi +\beta $. In this way,
\ we get to define the regularized generating functional as being
\begin{eqnarray}
Z^{^{\Lambda }}[\eta ,\bar{\eta}] &=&\!\!\int \!\!\mathcal{D}\rho \mathcal{D}%
\phi \,\mathcal{D}\psi \mathcal{D}\bar{\psi}\,\mathcal{D}\alpha \mathcal{D}%
\beta \;\exp \left( i\int \!\!dx-\frac{b}{2\pi }\;\partial _{\mu }\phi
\partial ^{\mu }\phi +\frac{b}{2\pi }\left[ \partial _{\mu }\beta \partial
^{\mu }\beta -\Lambda _{_{\beta }}^{2}\beta ^{2}\right] \right)
\label{th-eqx48} \\
&&\hspace{3.37cm}\;\exp \left( i\int \!\!dx\;\frac{b-1}{2\pi }\;\partial
_{\mu }\rho \partial ^{\mu }\rho -\frac{b-1}{2\pi }\left[ \partial _{\mu
}\alpha \partial ^{\mu }\alpha -\Lambda _{_{\alpha }}^{2}\alpha ^{2}\right]
\right)   \nonumber \\
&&\hspace{3.37cm}\;\exp \left( \,i\!\int \!\!dx\,\bar{\psi}i\partial \!\!\!%
\slash\psi +\bar{\eta}e^{i\left[ \rho +\alpha -\gamma _{5}(\phi +\beta )%
\right] }\psi +\bar{\psi}e^{-i\left[ \rho +\alpha +\gamma _{5}(\phi +\beta )%
\right] }\eta \right) .  \nonumber
\end{eqnarray}%
Now, before carrying out the integration of $\alpha $ and $\beta $ fields,
we make a change of variables $\rho \rightarrow \rho -\alpha $ and $\phi
\rightarrow \phi -\beta $. Thus, we obtain the following expression for the
regularized generating functional
\begin{eqnarray}
Z^{^{\Lambda }}[\eta ,\bar{\eta}] &=&\int \!\!\mathcal{D}%
\rho \mathcal{D}\phi \,\mathcal{D}\psi \mathcal{D}\bar{\psi}\;\exp \!\left(
i\int \!\!dx\;\bar{\psi}{i}\partial \!\!\!\slash\psi +\bar{\eta}e^{i\rho
-i\gamma _{5}\phi }\psi +\bar{\psi}{e}^{-i\rho -i\gamma _{5}\phi }\eta
\right)   \label{th25} \\
&&\hspace{2.6cm}\exp \left( i\int \!\!dx\;\frac{b}{2\pi \Lambda
_{_{\beta
}}^{2}}\;\phi \square (\square +\Lambda _{_{\beta }}^{2})\phi -\frac{b-1}{%
2\pi \Lambda _{_{\alpha }}^{2}}\;\rho \square (\square +\Lambda _{_{\alpha
}}^{2})\rho \right) .  \nonumber
\end{eqnarray}%
In this way we were able to improve the ultraviolet behavior of both
propagators. Being able to regularize those propagators, we are going to
express the regularized generating functional in function of the original
fields as in (\ref{th-eqx43}). Doing the change of fermionic variables
inverse to that one made in (\ref{th22}), and expressing fields $\rho $ and $%
\phi $ as functions of the field $A_{\mu }$ by using (\ref{th-eqx40}), we
obtain
\begin{equation}
Z^{^{\Lambda }}[\eta ,\bar{\eta}]=\!\int \!\!\mathcal{D}A_{\mu }\mathcal{D}%
\psi \mathcal{D}\bar{\psi}\;\exp \left( i\!\int \!\!dx\;\bar{\psi}(i\partial
\!\!\!\slash+A\!\!\!\slash)\psi +\frac{1}{2g}A_{\mu }A^{\mu }+\frac{b}{2\pi
\Lambda _{_{\beta }}^{2}}\;(\tilde{\partial}\cdotp A)^{2}-\frac{b-1}{2\pi
\Lambda _{_{\alpha }}^{2}}\;(\partial \cdotp A)^{2}+\bar{\eta}\psi +\bar{\psi%
}\eta \right)   \label{th26}
\end{equation}%
Comparing the Lagrangian densities that appear in the generating
functionals (\ref{th-eqx43}) and (\ref{th26}), we can see that the
regularization process has given rising to two new terms which
depend explicitly on the Pauli-Villars cut-offs, such terms are
enough to turn completely finite the model. We can to compute all
the regularized fermionic Green's functions starting from
(\ref{th25}) or (\ref{th26}).

By integrating the auxiliary field $A_{\mu }$, we get the regularized
generating functional in terms of the fermion fields alone,
\begin{equation}
Z^{^{\Lambda }}[\eta ,\bar{\eta}]\;=\!\int \!\!\mathcal{D}\psi \mathcal{D}%
\bar{\psi}\,\exp \left( i\!\int \!\!dx\;\bar{\psi}i\partial \!\!\!\slash\psi
-\frac{g}{2}\;(\bar{\psi}\gamma ^{\mu }\psi )^{2}+\frac{g}{2}\;\bar{\psi}%
\gamma ^{\mu }\psi D_{\mu \nu }^{^{\Lambda }}\bar{\psi}\gamma ^{\nu }\psi +%
\bar{\eta}\psi +\bar{\psi}\eta \right) \!,  \label{th-eqx51}
\end{equation}%
where the operator $D_{\mu \nu }^{^{\Lambda }}$ carries all the information
about the non--local counter-terms which regularize the fermionic theory,
\begin{equation}
D_{\mu \nu }^{^{\Lambda }}\;=\;\frac{\partial _{\mu }\partial _{\nu }}{%
\square +{\displaystyle\frac{\pi \Lambda _{_{\alpha }}^{2}}{g(b-1)}}}\;+\;%
\frac{g_{\mu \nu }\square -\partial _{\mu }\partial _{\nu }}{\square +{%
\displaystyle\frac{\pi \Lambda _{_{\beta }}^{2}}{gb}}}\,,  \label{th29}
\end{equation}%
as we can see, it shows explicitly the longitudinal and transversal
contributions that regularize the respective components of the fermionic
current--current interaction.

%%%%%%%%%%%%%%%%%%%%%%%%%%%%%%%%%%%%%%%%%%%%%%%%%%%%%%%%%%%%%%%%%%%%%

\subsubsection{Regularized and renormalized fermion propagator \label%
{propregren}}

\bigskip

We start from the regularized generating functional (\ref{th25}),
integrating over the fermionic fields it reads as
\begin{equation}
Z^{^{\Lambda }}[\eta ,\bar{\eta}]=\!\!\int \!\!\mathcal{D}\rho \mathcal{D}%
\phi \;\exp \left( i\int \!\!dx\;\frac{b}{2\pi \Lambda _{_{\beta }}^{2}}%
\;\phi \square (\square +\Lambda _{_{\beta }}^{2})\phi -\frac{b-1}{2\pi
\Lambda _{_{\alpha }}^{2}}\;\rho \square (\square +\Lambda _{_{\alpha
}}^{2})\rho \right) \exp \left( -i\!\int \!\!dx\,dy\;\bar{\eta}(x)G(x,y;\rho
,\phi )\eta (y)\right) ,  \label{th30}
\end{equation}%
where the Green's function $G(x,y;\rho ,\phi )$ is given by (\ref{th-eqx42}%
). Then, the regularized fermion propagator in configuration space is given
by
\begin{equation}
G^{^{\Lambda }}(x-y)=i\;\exp \left( i\int \!\!\frac{dk}{(2\pi )^{2}}\left[ \;%
\frac{\pi \Lambda _{_{\alpha }}^{2}}{b-1}\;\frac{1}{k^{2}(k^{2}-\Lambda
_{_{\alpha }}^{2})}-\frac{\pi \Lambda _{_{\beta }}^{2}}{b}\;\frac{1}{%
k^{2}(k^{2}-\Lambda _{_{\beta }}^{2})}\right] \left[ 1-e^{-ik\cdotp(x-y)}%
\right] \right) G_{F}(x-y)\;,
\end{equation}%
performing the momentum integration in the exponential and introducing the
arbitrary mass scale $M$, the propagator is
\begin{equation}
G^{^{\Lambda }}(x,y)=\frac{1}{2\pi i}\frac{x\!\!\!\slash-y\!\!\!\slash}{%
(x-y)^{2}-i\varepsilon }\left( \frac{\tilde{\Lambda}_{\alpha }^{2}}{M^{2}}%
\right) ^{\displaystyle-\frac{1}{4(b-1)}}\left( \frac{\tilde{\Lambda}_{\beta
}^{2}}{M^{2}}\right) ^{\displaystyle\frac{1}{4b}}\left[ \frac{{}}{{}}%
\!\!-M^{2}(x-y)^{2}+i\varepsilon \right] ^{\displaystyle-\frac{1}{c}}
\label{th-eqx55}
\end{equation}%
where the parameter $c$ is defined as being
\begin{equation}
c\;=\;4b(b-1)\;=\;\left( \frac{2\pi }{g}+a\right) ^{2}-1\,,  \label{th34x}
\end{equation}%
we can see that $c\in \;[-1,+\infty >$ and $c^{-1}\in \;<-\infty ,-1]\;\cup
<0,+\infty >$.

In this way, we can renormalize exactly in configuration space
\cite{Helayeldica}.The renormalization procedure establishes the
following relation between renormalized and bare (regularized) 1PI
Green's function, $\Gamma^{^R}\!=Z_\psi\, \Gamma^{^\Lambda}$ , where
$Z_\psi$ is the fermionic wave function renormalization constant
that relates the bare and renormalized fermionic fields, $%
\psi_{bare}=Z_\psi\psi_R$. Thus, from (\ref{th-eqx55}), we choose
the fermionic wave function renormalization constant to be
\begin{equation}
Z_{\psi }=\left( \frac{\tilde{\Lambda}_{\alpha }^{2}}{M^{2}}\right) ^{%
\displaystyle-\frac{1}{4(b-1)}}\left( \frac{\tilde{\Lambda}_{\beta }^{2}}{%
M^{2}}\right) ^{\displaystyle\frac{1}{4b}}  \label{th37-1}.
\end{equation}%
It can be shown that a renormalization of the fermionic wave function is
sufficient to get renormalize all the Green's functions of the model.

Then, the renormalized propagator of massless Thirring model in
configuration space is read as
\begin{equation}
G^{^{R}}(x-y)=\frac{1}{i\,2\pi }\frac{x\!\!\!\slash-y\!\!\!\slash}{%
(x-y)^{2}-i\epsilon }\left( \!\frac{{}}{{}}-M^{2}(x-y)^{2}+i\epsilon \right)
^{-1/c},  \label{th38-1}
\end{equation}%
in the limit $g\rightarrow 0$ we also obtain the propagator of a free
massless fermion field.

Then, by computing the Fourier transform we get the renormalized fermion
propagator in momentum space,
\begin{equation}
G^{^{R}}(p)=\frac{i}{p\!\!\!\slash}\left( \frac{\bar{M}^{2}}{p^{2}}\right)
^{\!\!-\displaystyle\frac{1}{c}}{e}^{-i\pi /c}\frac{\Gamma (1-{1}/{c})}{%
\Gamma (1+{1}/{c})}  \label{th-eqx59}
\end{equation}%
where $\overline{M}=2M$. The dynamical or anomalous dimension of the
massless Thirring fermion field is $1/c$. By choosing an adequate
value for the ambiguity parameter we can reproduce results such as
has been obtained in \cite{thirring1,thirring2,vglaser1,shroer1}.
This value is got when we set $c=2\pi /g$, thus, the ambiguity
parameter which reproduces the result to the anomalous dimension
found in the literature is
\begin{equation}
a_{_{l}}:=-\frac{2\pi }{g}\pm \sqrt{1+\frac{2\pi }{g}}\;.
\end{equation}

In this value we will obtain a result which agrees well with Glaser's
analysis of the massless Thirring model \cite{thirring2,vglaser1} and
spectral analysis of two--point correlation functions in 1+1--dimensional
quantum field theories developed by Schroer \cite{shroer1}.

The renormalized propagator (\ref{th-eqx59}) has singularities in ${%
\displaystyle\frac{1}{c}} = \pm 1,\;\pm 2,\;\ldots$. And, if we set $%
c=2\pi/g $ the propagator (\ref{th-eqx59}) has singularities in $g=\pm 2\pi$
such as it was shown by Klaiber analysis \cite{klaiber1}.

%%%%%%%%%%%%%%%%%%%%%%%%%%%%%%%%%%%%%%%%%%%%%%%%%%%%%%%%%%%%%%%%%%%%%
%%%%%%%%%%%%%%%%%%%%%%%%%%%%%%%%%%%%%%%%%%%%%%%%%%%%%%%%%%%%%%%%%%%%%
%%%%%%%%%%%%%%%%%%%%%%%%%%%%%%%%%%%%%%%%%%%%%%%%%%%%%%%%%%%%%%%%%%%%%

\section{The gauge invariant case: $a=1-\frac{2 \protect\pi}{g}$}

%%%%%%%%%%%%%%%%%%%%%%%%%%%%%%%%%%%%%%%%%%%%%%%%%%%%%%%%%%%%%%%%%%%%%
%%%%%%%%%%%%%%%%%%%%%%%%%%%%%%%%%%%%%%%%%%%%%%%%%%%%%%%%%%%%%%%%%%%%%

As we have observed in Section 2, the Ward identity for the 1PI
2-point function of the fermionic current reveals the existence of a
local $U(1)$ gauge symmetry for $a=1-\frac{2\pi }{g}$. For this
value  such symmetry is explicit in the effective action for the
vector field in the generating functional (\ref{th-eqx7}), i.e., we
can not to perform the functional integration. Therefore, we need to
construct a well-defined generating functional for this gauge
theory. Such construction is making via the Faddeev-Popov procedure
\cite{faddeev} with Lorentz gauge fixing condition $\;f(A_{\mu
})={\displaystyle \frac{1}{\sqrt{\xi }}}\;\partial \cdotp A\;$  in
the generating functional (\ref{th-eqx4}). Thus, the resulting
generating functional for this gauge invariant theory reads as
\begin{equation}
Z^{^{\xi }}[\eta ,\bar{\eta},C_{\mu }]=\!\!\int \!\!\mathcal{D}A_{\mu }%
\mathcal{D}\psi \mathcal{D}\bar{\psi}\;\exp \left( i\!\int \!\!dx\;\bar{\psi}%
(i\partial \!\!\!\slash+A\!\!\!\slash+C\!\!\!\!\slash)\psi +\frac{1}{2g}%
\;A_{\mu }A^{\mu }-\frac{1}{2\xi }\;(\partial \cdotp A)^{2}+\bar{\eta}\psi +%
\bar{\psi}\eta \right) ,  \label{th-eqx61}
\end{equation}%
where we have introduce the external source $C_{\mu }$ for the
fermionic current. From this generating functional we can compute
all the Green's functions and we can also to construct the
Schwinger--Dyson equations (SDE) and the Ward identities.

%%%%%%%%%%%%%%%%%%%%%%%%%%%%%%%%%%%%%%%%%%%%%%%%%%%%%%%%%%%%%%%%%%%%%

\subsection{The Green's functions}

%%%%%%%%%%%%%%%%%%%%%%%%%%%%%%%%%%%%%%%%%%%%%%%%%%%%%%%%%%%%%%%%%%%%%

\subsubsection{The fermion propagator $G^{^{\xi}}(x-y)=\langle 0|T\psi(x)
\bar{\psi}(y)|0\rangle $}

%%%%%%%%%%%%%%%%%%%%%%%%%%%%%%%%%%%%%%%%%%%%%%%%%%%%%%%%%%%%%%%%%%%%%

We set $C_{\mu }=0$ in the generating functional (\ref{th-eqx61}) and
integrating the fermionic degrees of freedom we obtain
\begin{equation}
Z^{^{\xi }}[\eta ,\bar{\eta}]=\!\int \!\!\mathcal{D}A_{\mu }\;\exp \left(
i\!\int \!\!dx\;\frac{1}{2}A_{\mu }\left[ \frac{1}{\pi }\;g^{\mu \nu }-\frac{%
1}{\pi }\;\frac{\partial ^{\mu }\partial ^{\nu }}{\square }+\frac{1}{\xi }%
\;\partial ^{\mu }\partial ^{\nu }\right] A_{\mu }\right) \exp \left(
-i\!\int \!\!dx\,dy\;\bar{\eta}(x)G(x,y;A)\eta (y)\right) ,  \label{th-eqx62}
\end{equation}%
where the Green's function $G(x,y;A)$ is given in (\ref{th-eqx9}).
From this, we calculate the fermion propagator which becomes
\begin{equation}
G^{^{\xi }}(x-y)=i\;\exp \left( i\int \!\!\frac{dk}{(2\pi )^{2}}\left( \frac{%
\xi }{k^{4}}\;+\;\frac{\pi }{k^{2}}\right) \left[ 1-e^{-ik\cdotp(x-y)}\right]
\right) G_{F}(x-y)\;,  \label{th-eqx63}
\end{equation}%
We can see the existence of logarithmic UV divergences coming from
the term in the integral proportional to $k^{-2}$ and,  we also
observe that the term proportional to the gauge parameter $\xi $
generates serious infrared divergences which can not be
renormalized. These infrared divergences proportional to $\xi $ are
also observed in the all Green's functions computed from the
generating functional (\ref{th-eqx61}), then, we use a gauge free of
infrared divergences $\;\xi =0$.

%%%%%%%%%%%%%%%%%%%%%%%%%%%%%%%%%%%%%%%%%%%%%%%%%%%%%%%%%%%%%%%%%%%%%

\subsubsection{The current propagator $J_{\mu\nu}^{^{\xi}}=\langle 0|T
j_{\mu}(x)j_{\nu}(y)|0\rangle $}

%%%%%%%%%%%%%%%%%%%%%%%%%%%%%%%%%%%%%%%%%%%%%%%%%%%%%%%%%%%%%%%%%%%%%

In momentum space the propagator of the fermionic current is calculated to
be
\begin{equation}
i\,J_{\mu \nu }^{^{^{\xi }}}(q)=\frac{1}{g}\left( 1-\frac{\pi }{g}\right)
T_{\mu \nu }+\frac{1}{g}\left( \frac{\xi }{gk^{2}}-1\right) L_{\mu \nu }
\label{th-eqx64}
\end{equation}%
it is free of UV divergences.

%%%%%%%%%%%%%%%%%%%%%%%%%%%%%%%%%%%%%%%%%%%%%%%%%%%%%%%%%%%%%%%%%%%

\subsubsection{The vertex function $G_{\mu }^{^{\xi}}(x,y;z)=\langle 0|T
\psi(x)\bar{\psi}(y)j_{\mu}(z)|0\rangle $}

%%%%%%%%%%%%%%%%%%%%%%%%%%%%%%%%%%%%%%%%%%%%%%%%%%%%%%%%%%%%%%%%%%%

The vertex function is calculated to be
\begin{equation}
{G}_{\mu }^{^{\xi }}(x,y;z)=i\int \!\!\frac{dk}{(2\pi )^{2}}\left( (\frac{%
\xi }{gk^{2}}-1)\frac{k_{\mu }}{k^{2}}+\frac{\pi }{g}\frac{\tilde{k}_{\mu }}{%
k^{2}}\,\gamma _{5}\right) \left( e^{-ik\cdotp(z-x)}-e^{-ik\cdotp%
(z-y)}\right) G^{\xi }(x-y),  \label{th-eqx65}
\end{equation}%
in momentum space it reads as
\begin{equation}
G_{\mu }^{^{\xi }}(p,-p-q;q)=i\left[ \left( \frac{\xi }{gq^{2}}-1\right)
\frac{q_{\mu }}{q^{2}}+\frac{\pi }{g}\frac{\tilde{q}_{\mu }}{q^{2}}\,\gamma
_{5}\right] \left[ \,G^{^{\xi }}(p+q)-G^{^{\xi }}(p)\frac{{}}{{}}\!\right] .
\label{th-eqx66}
\end{equation}%
where $G^{^{\xi }}(p)$ is the Fourier transforms of the fermionic propagator
(\ref{th-eqx63}). Thus, we can note that the UV divergences of this Green`s
function come from the fermionic propagator.

%%%%%%%%%%%%%%%%%%%%%%%%%%%%%%%%%%%%%%%%%%%%%%%%%%%%%%%%%%%%%%%%%%%%%

\subsubsection{The 1PI functions}

%%%%%%%%%%%%%%%%%%%%%%%%%%%%%%%%%%%%%%%%%%%%%%%%%%%%%%%%%%%%%%%%%%%%%

In momentum space the 1PI 2-point fermion function $\Gamma ^{^{\xi }}(p)$ is
\begin{equation}
\Gamma ^{^{\xi }}(p)=i\left[ G^{^{\xi }}(p)\right] ^{-1}.  \label{th-eqx67}
\end{equation}

Also, the 1PI 2-point function for the fermionic current $\Gamma _{\mu \nu
}^{^{\xi }}(q)$ is%
\begin{equation}
\Gamma _{\mu \nu }^{^{\xi }}(q)=-{\displaystyle{\frac{g}{1-\displaystyle%
\frac{\pi }{g}}}}\;T_{\mu \nu }-\displaystyle\frac{g}{\displaystyle{\frac{%
\xi }{gq^{2}}-1}}\;L_{\mu \nu }  \label{th-eqx68-a}
\end{equation}

The 1PI 3-point function, $\Gamma _{\mu }^{^{\xi }}(p,k;q)$ related to the
vertex function $G_{\mu }^{^{\xi }}(p,k;q)$ is given by
\begin{equation}
\Gamma _{\mu }^{^{\xi }}(p,-p-q;q)=\left( -g\,\frac{q_{\mu }}{q^{2}}+\frac{%
\pi }{1-\displaystyle\frac{\pi }{g}}\,\frac{\tilde{q}_{\mu }}{q^{2}}\;\gamma
_{5}\right) \left[ \Gamma ^{^{\xi }}(p+q)\frac{{}}{{}}\!\!\,-\Gamma ^{^{\xi
}}(p)\right]  \label{th-eqx69}
\end{equation}

%%%%%%%%%%%%%%%%%%%%%%%%%%%%%%%%%%%%%%%%%%%%%%%%%%%%%%%%%%%%%%%%%%%

\subsection{The Schwinger-Dyson equations}

%%%%%%%%%%%%%%%%%%%%%%%%%%%%%%%%%%%%%%%%%%%%%%%%%%%%%%%%%%%%%%%%%%%

We start from the well-defined gauge fixed generating functional
(\ref{th-eqx61}) which after the integration of the vector field reads
as
\begin{equation}
Z^{^{\xi }}[\eta ,\bar{\eta},C_{\mu }]=\!\int \!\!\mathcal{D}\psi \mathcal{D}%
\bar{\psi}\;\exp \left( i\!\int \!\!dx\;\bar{\psi}(i\partial \!\!\!\slash%
+C\!\!\!\!\slash)\psi -\frac{g}{2}\,j_{\mu }j^{\mu }+\frac{g}{2}\,j^{\mu
}D_{\mu \nu }^{\xi }j^{\nu }+\bar{\eta}\psi +\bar{\psi}\eta \right) \;,
\label{th-eqx71}
\end{equation}%
where the third term appears due to the gauge fixing condition, and the
operator $D_{\mu \nu }^{\xi }$ is given as
\begin{equation}
D_{\mu \nu }^{^{\xi }}=\frac{\partial _{\mu }\partial _{\nu }}{\square +%
\displaystyle\frac{\xi }{g}}\,\;.  \label{th-eqx72}
\end{equation}

By using the functional representation for the fields (\ref{th-eqq23}) in
the equation of motion for the fermionic field, we get the quantum equation
of motion for the generating functional (\ref{th-eqx71})
\begin{equation}
\left( \partial _{x}\!\!\!\!\!\!\slash\;\;\frac{\delta }{\delta \bar{\eta}(x)%
}+g\gamma ^{\mu }\frac{\delta }{\delta \bar{\eta}(x)}\frac{\delta }{\delta
C^{\mu }(x)}-iC_{\mu }(x)\gamma ^{\mu }\frac{\delta }{\delta \bar{\eta}(x)}%
+\eta (x)-g\gamma ^{\mu }\;\frac{\delta }{\delta \bar{\eta}(x)}D_{\mu \nu
}^{^{\xi }}\frac{\delta }{\delta C_{\nu }(x)}\right) Z^{^{\xi }}[\eta ,\bar{%
\eta},C^{\mu }]=0  \label{th-eqx73}
\end{equation}%
The last term makes a difference between the SDE's (\ref{th-eqx73}) and (\ref%
{th-eqx22}) but it arises due to the introduction of a gauge fixing term
into the generating functional (\ref{th-eqx61}) for well-defined it. Thus,
in a formal treatment the gauge fixing term is not present and therefore the
SDE for the gauge invariant phase is exactly the same as in (\ref{th-eqx22})
for the non-gauge invariant phase .

Thus, the SDE for the fermion propagator in the gauge invariant phase is
given by
\begin{equation}
i\partial \!\!\!\slash_{x}G^{^{\xi }}(x-y)\;=\;i\delta (x-y)+g\gamma ^{\mu
}G_{\mu }^{^{\xi }}(x,y;x)-g\gamma ^{\mu }\lim_{z\rightarrow x}D_{\!\!z\;\mu
\nu }^{^{\xi }}G_{\mu }^{\xi }(x,y;z)
\end{equation}%
Both the fermionic propagator (\ref{th-eqx63}) and the vertex function (\ref%
{th-eqx65}) computed from the generating functional (\ref{th-eqx61}) are
also solutions for the quantum equation of motion of the massless Thirring
model.

In momentum space the SDE for the fermion propagator reads as
\begin{equation}
G^{^{\xi }}(p)=\frac{i}{p\!\!\!\slash}\;-\;i\int \!\!\frac{dk}{(2\pi )^{2}}%
\;f^{^{\xi }}(k)\;\frac{1}{p\!\!\!\slash}\;k\!\!\!\slash\;G^{^{\xi }}(p-k)\;,
\label{th-eqx75}
\end{equation}%
where the function $f^{^{\xi }}(k)$ reads for
\begin{equation}
f^{^{\xi }}(k)=\frac{\xi }{k^{4}}+\frac{\pi }{k^{2}}\;.  \label{th-eqx76}
\end{equation}

Similarly to the case \textit{non--gauge invariant} (\ref{th14x}) by
iterating the SDE it generates an expansion in $\hbar $ or
equivalently in powers of $f^{^{\xi }}(k)$. The power counting
analysis of the series shows clearly the existence of the same
logarithmic UV divergences in each order or loop what indicates that
the model is a fully renormalizable field theory.

%%%%%%%%%%%%%%%%%%%%%%%%%%%%%%%%%%%%%%%%%%%%%%%%%%%%%%%%%%%%%%%%%%%%%

\subsection{The Ward identities}

%%%%%%%%%%%%%%%%%%%%%%%%%%%%%%%%%%%%%%%%%%%%%%%%%%%%%%%%%%%%%%%%%%%%%

We start from the well-defined generating functional (\ref{th-eqx61}) and we
make the following fermionic transformation
\begin{equation}
\psi \rightarrow e^{i\lambda (x)}\psi \quad ,\quad \bar{\psi}\rightarrow
\bar{\psi}e^{-i\lambda (x)}\,,  \label{th-eqx78}
\end{equation}%
one more time it is worthwhile to note that in our framework the fermionic
measure is not invariant under this transformation changing as $\mathcal{D}%
\psi \mathcal{D}\bar{\psi}\rightarrow J^{^{\xi }}[A,\lambda]\,
\mathcal{D}\psi \mathcal{D}\bar{\psi}$ where $J^{^{\xi }}[A,\lambda]$
is the Jacobian given by
\begin{equation}
J^{^{\xi }}[A,\lambda]\;=\exp \left( i\int
\!\!dx\,-\frac{1}{2g}\;(A_{\mu }+C_{\mu })\partial ^{\mu }\lambda
+\frac{1}{2g}\;(\partial _{\mu }\lambda )^{2}\right) .
\label{th-eqx79}
\end{equation}%
The generating functional (\ref{th-eqx61}) after the fermionic
transformation reads as
\begin{equation}
Z^{^{\xi }}[\eta ,\bar{\eta},C_{\mu }]=\!\int \!\!\mathcal{D}A_{\mu }%
\mathcal{D}\psi \mathcal{D}\bar{\psi}J^{^{\xi }}[A]\,\exp \left( i\!\int
\!\!dx\;\bar{\psi}\left( {i}\partial +A\!\!\!\slash+C\!\!\!\!\slash-\partial
\!\!\!\slash\lambda \right) \psi +\frac{1}{2g}\,A\cdotp{\!\!A}-\frac{1}{2\xi }%
(\partial \cdotp {\!\!A})^{2}+\bar{\eta}e^{i\lambda (x)}\psi +\bar{\psi}%
e^{-i\lambda (x)}\eta \right) .  \label{th-eqx80}
\end{equation}%
At once we make the transformation $A_{\mu }\rightarrow A_{\mu }+\partial
_{\mu }\lambda $ and after integrating over the vector field we obtain
\begin{eqnarray}
Z^{^{\xi }}[\eta ,\bar{\eta},C_{\mu }] &=&\!\!\int \!\!\mathcal{D}\psi
\mathcal{D}\bar{\psi}\;\exp \left( i\int \!\!dx\;\bar{\psi}({i}\partial
\!\!\!\slash+C\!\!\!\!\slash\,)\psi -\frac{g}{2}\;(\bar{\psi}\gamma ^{\mu
}\psi )^{2}+\frac{g}{2}\;j^{\mu }D_{\mu \nu }^{^{\xi }}j^{\nu }+\bar{\eta}%
e^{i\lambda (x)}\psi +\bar{\psi}e^{-i\lambda (x)}\eta \right)
\label{th-eqx81} \\
&&\hspace{1.4cm}\exp \left( i\int \!\!dx-\frac{1}{g}\,C_{\mu }\partial ^{\mu
}\lambda +\lambda \frac{\square }{\square +\displaystyle\frac{\xi }{g}}%
\;\partial _{\mu }j^{\mu }-\frac{1}{2g}\;\lambda \frac{\square ^{2}}{\square
+\displaystyle\frac{\xi }{g}}\;\lambda \right)  \nonumber
\end{eqnarray}

Considering $\lambda (x)$ as an infinitesimal function and expanding the
exponential to first order in $\lambda $ we get the functional equation
satisfied by the generating functional (\ref{th-eqx61}),
\begin{equation}
\left( \bar{\eta}\,\frac{\delta }{\delta \bar{\eta}}-\eta \,\frac{\delta }{%
\delta \eta }+\frac{1}{g}\,\partial _{\mu }{C}^{\mu }-i\,\frac{\square }{%
\square +\displaystyle\frac{\xi }{g}}\,\partial _{\mu }\frac{\delta }{\delta
{C}_{\mu }}\right) Z^{^{\xi }}[\eta ,\bar{\eta},C_{\mu }]=0  \label{th-eqx82}
\end{equation}%
where we have used the functional representations for the fields (\ref%
{th-eqq23}).

We define the generating functional of the connected Green's
functions $W^{^{\xi }}[\eta ,\bar{\eta},C_{\mu }]$  as being
$W^{^{\xi }}=-i\ln Z^{^{\xi }}$ and the generating functional of the
1PI functions $\Gamma ^{^{\xi }}[\bar{\psi},\psi ,j^{\mu }]$
 which are related via a Legendre transformation similar to
(\ref{th-eqx30}). Using the functional relations given in
(\ref{th-eqx31}), we obtain the functional equation satisfied by the
generating functional $\Gamma ^{^{\xi}}[\bar{\psi},\psi ,j^{\mu }]$
from which we get all Ward identities
\begin{equation}
i\,\frac{\delta \Gamma ^{^{\xi }}}{\partial \psi (x)}\,\psi (x)-i\,\frac{%
\delta \Gamma ^{^{\xi }}}{\partial \bar{\psi}(x)}\bar{\psi}(x)-\frac{1}{g}%
\,\partial _{\mu }^{x}\,\frac{\delta \Gamma ^{^{\xi }}}{\partial {j}_{\mu
}(x)}+\frac{\square _{x}}{\square _{x}+\displaystyle\frac{\xi }{g}}%
\;\partial _{\mu }^{x}\,{j}^{\mu }(x)=0  \label{th-eqx85}.
\end{equation}%
If we observe the same functional equation in the \textit{non--gauge
invariant phase} (\ref{th-eqx32}), the equation above can be
obtained by setting $a=1-\displaystyle{\frac{2\pi }{g}}\,$. The last
term appears due to the gauge fixing condition used to define the
generating functional of the Green's functions in this phase.

\vskip 0.5cm

The first Ward identity is to the 1PI 2-point function $\Gamma _{\mu \nu
}^{^{\xi }}(q)$ for the fermionic current which reads as
\begin{equation}
q^{\mu }\Gamma _{\mu \nu }^{^{\xi }}(q)=g\;\frac{q^{2}}{q^{2}-\displaystyle%
\frac{\xi }{g}}\;q^{\nu }\;  \label{th-eqx86}
\end{equation}%
as it can be shown explicitly from the equation (\ref{th-eqx68-a}). Thus,
the right-handed side above is gauge dependent what guarantees that the
current propagator is transverse for the critical value $a=1-\displaystyle{%
\frac{2\pi }{g}}\,$, as it was noted in (\ref{th-eqx33-a}).

The second Ward identity relates the 1PI $3$-point function $\Gamma _{\mu
}^{^{\xi }}(p;q)$ and the 1PI 2-point fermion function $\Gamma ^{^{\xi }}(p)$
is given by
\begin{equation}
q^{\mu }\Gamma _{\mu }^{^{\xi }}(p;q)=-g\left[ \Gamma ^{^{\xi }}(p+q)-\Gamma
^{^{\xi }}(p)\right]  \label{th-eqx87}
\end{equation}%
likewise it can also be shown easily from the equation (\ref{th-eqx69}).
This identity shows that the coupling constant does not renormalize due to a
renormalization of the fermionic propagator guarantees the renormalization
of this 1PI $3$-point function.

%%%%%%%%%%%%%%%%%%%%%%%%%%%%%%%%%%%%%%%%%%%%%%%%%%%%%%%%%%%%%%%%%%%%%

\subsection{Regularization and renormalization of the UV divergences}

%%%%%%%%%%%%%%%%%%%%%%%%%%%%%%%%%%%%%%%%%%%%%%%%%%%%%%%%%%%%%%%%%%%%%

We will study the divergences of the pure fermionic Green's functions
(Green's functions without current insertions) following the procedure
achieved in the subsection \ref{regbneq1} for the \textit{non--gauge
invariant phase}. Therefore, starting from the generating functional (\ref%
{th-eqx62}) and decomposing the vectorial field $A_{\mu }$ in their
longitudinal and transversal parts (\ref{th-eqx40}) we arrive to the
following expression for the generating functional for the all pure
fermionic Green's functions
\begin{equation}
Z^{^{\xi }}[\eta ,\bar{\eta}]=\!\int \!\!\mathcal{D}\rho \mathcal{D}\phi
\;\exp \left( i\!\int \!\!dx\;-\frac{1}{2\pi }\;\partial _{\mu }\phi
\partial ^{\mu }\phi -\frac{1}{2\xi }\;(\square \rho )^{2}\right) \exp
\left( -i\!\int \!\!dx\,dy\;\bar{\eta}(x)G(x,y;\rho ,\phi )\eta (y)\right) ,
\label{thb1-7}
\end{equation}%
where the Green's function $G(x,y;\rho ,\phi )$ is given by equation (\ref%
{th-eqx42}). Thus, in order to compute the fermionic propagator we
take functional derivatives with respect to the external fermionic
sources and setting $\eta =\bar{\eta}=0$ we obtain
\begin{equation}
G^{^{\xi }}(x-y)=i\!\int \!\!\mathcal{D}\rho \mathcal{D}\phi \;\exp \left(
i\!\int \!\!dx\;-\frac{1}{2\pi }\;\partial _{\mu }\phi \partial ^{\mu }\phi -%
\frac{1}{2\xi }\;(\square \rho )^{2}\right) G(x,y;\rho ,\phi )\,,
\end{equation}%
where the integrations over the field $\rho $ and $\phi $ are
factorized. It can easily be observed that the UV divergences come
from the integration on the transversal field $\phi $ due to the bad
behavior of its propagator at high energy, thus, we need to improve
its behavior at high energy to be able to control the UV
divergences.

Therefore, we start the regularization process from the generating
functional (\ref{th-eqx61}) setting $C_{\mu }=0$, then, we decompose the
vector field as in (\ref{th-eqx40}) and make the fermionic transformation (%
\ref{th22}) which changes the fermionic measure as $\mathcal{D}\psi \mathcal{%
D}\bar{\psi}\;\rightarrow \,J^{^{\xi }}[\rho ,\phi ]\mathcal{D}\psi \mathcal{%
D}\bar{\psi}$, where $J^{^{\xi }}[\rho ,\phi ]$ is the Jacobian given by
\begin{equation}
J^{^{\xi }}[\rho ,\phi ]=\exp \left( i\!\int \!\!dx\;-\frac{1}{2g}\;\partial
_{\mu }\rho \partial ^{\mu }\rho -\frac{1}{2g}\left( \frac{g}{\pi }-1\right)
\partial _{\mu }\phi \partial ^{\mu }\phi \right) \,.
\end{equation}%
Finally, we get the following expression for the generating functional
\begin{equation}
Z^{^{\xi }}[\eta ,\bar{\eta}]=\!\!\int \!\!\mathcal{D}\rho \mathcal{D}\phi \,%
\mathcal{D}\psi \mathcal{D}\bar{\psi}\;\exp \!\left( i\!\int \!\!dx\ \bar{%
\psi}\,i\partial \!\!\!\slash\,\psi -\frac{1}{2\xi }\;(\square \rho )^{2}-%
\frac{1}{2\pi }\;\partial _{\mu }\phi \partial ^{\mu }\phi +\bar{\eta}%
e^{i\rho -i\gamma _{5}\phi }\psi +\bar{\psi}e^{-i\rho -i\gamma _{5}\phi
}\eta \right)
\end{equation}%
Then, we following the regularization procedure described between the
equations (\ref{th-eqx47}) and (\ref{th25}) -for the non--gauge invariant
phase- but improving only the $\phi $--propagator. Such procedure leads to
the following regularized generating functional for the gauge invariant
phase
\begin{equation}
Z^{^{\Lambda \xi }}[\eta ,\bar{\eta}]=\!\!\int \!\!\mathcal{D}\rho \mathcal{D%
}\phi \,\mathcal{D}\psi \mathcal{D}\bar{\psi}\;\exp \left( i\!\int \!\!dx\;%
\bar{\psi}i\partial \!\!\!\slash\,\psi -\frac{1}{2\xi }\;(\square \rho )^{2}+%
\frac{1}{2\pi \Lambda _{_{\beta }}^{2}}\;\phi \square (\square +\Lambda
_{_{\beta }}^{2})\phi +\bar{\eta}e^{i\rho -i\gamma _{5}\phi }\psi +\bar{\psi}%
{e}^{-i\rho -i\gamma _{5}\phi }\eta \right) .  \label{th-eqx92}
\end{equation}%
Such regularized generating functional expressed in terms of the original
fields reads as
\begin{equation}
Z^{^{\Lambda \xi }}[\eta ,\bar{\eta}]=\!\!\int \!\!\mathcal{D}A_{\mu }%
\mathcal{D}\psi \mathcal{D}\bar{\psi}\;\exp \left( i\!\int \!\!dx\;\bar{\psi}%
(i\partial \!\!\!\slash+A\!\!\!\slash)\psi +\frac{1}{2g}A_{\mu }A^{\mu }+%
\frac{1}{2\pi \Lambda _{_{\beta }}^{2}}\;(\tilde{\partial}\cdotp A)^{2}-%
\frac{1}{2\xi }\;(\partial \cdotp A)^{2}+\bar{\eta}\psi +\bar{\psi}\eta
\right)  \label{th-eqx93}
\end{equation}%
or, we can also express it as a pure fermionic theory integrating
the auxiliary field $A_{\mu }$
\begin{equation}
Z^{^{\Lambda \xi }}[\eta ,\bar{\eta}]=\!\!\int \!\!\mathcal{D}\psi \mathcal{D%
}\bar{\psi}\exp \!\left( i\!\int \!\!dx\;\bar{\psi}i\partial \!\!\!\slash%
\psi -\frac{g}{2}\;(\bar{\psi}\gamma ^{\mu }\psi )^{2}+\frac{g}{2}\;\bar{\psi%
}\gamma ^{\mu }\psi D_{\mu \nu }^{^{\Lambda \xi }}\bar{\psi}\gamma ^{\nu
}\psi +\bar{\eta}\psi +\bar{\psi}\eta \right) ,  \label{th-eqx94}
\end{equation}%
where the operator $D_{\mu \nu }^{^{\Lambda \xi }}$ is defined as
\begin{equation}
D_{\mu \nu }^{^{\Lambda \xi }}\;=\;\frac{\partial _{\mu }\partial _{\nu }}{%
\;\square +{\displaystyle\frac{\xi }{g}}\;}\;+\;\frac{g_{\mu \nu }\square
-\partial _{\mu }\partial _{\nu }}{\square +{\displaystyle\frac{\pi \Lambda
_{_{\beta }}^{2}}{g}}}\;.  \label{thb1-14}
\end{equation}%
The third term in the regularized generating functional (\ref{th-eqx94}) has
two contributions: the longitudinal one which has already observed in (\ref%
{th-eqx71}) is due to the gauge fixing condition and the transversal
one that arises due to the regularization procedure of the UV
divergences. Thus, the gauge fixing condition acts as a natural
regulator of the longitudinal part of the current--current
interaction and the Pauli-Villars method was necessary to regularize
the transversal component of the interaction.

%%%%%%%%%%%%%%%%%%%%%%%%%%%%%%%%%%%%%%%%%%%%%%%%%%%%%%%%%%%%%%%%%%%

\subsubsection{Regularized and renormalized fermion propagator}

%%%%%%%%%%%%%%%%%%%%%%%%%%%%%%%%%%%%%%%%%%%%%%%%%%%%%%%%%%%%%%%%%%%

From the regularized generating functional (\ref{th-eqx92}) we can compute
all the regularized pure fermionic Green's functions of the local gauge
invariant model. Thus, for arbitrary gauge $\xi $ the regularized fermion
propagator is given by
\begin{equation}
G^{^{\Lambda \xi }}(x-y)=i\;\exp \left( i\int \!\!\frac{dk}{(2\pi )^{2}}\;%
\left[ \frac{\xi }{k^{4}}-\frac{\pi \Lambda _{_{\beta }}^{2}}{%
k^{2}(k^{2}-\Lambda _{_{\beta }}^{2})}\right] \left[ 1-e^{-ik\cdotp(x-y)}%
\right] \right) G_{F}(x-y)\;,  \label{th-eqx96}
\end{equation}%
Now, going to the $\xi =0$ gauge we compute the remaining integration to
obtain the following expression for the regularized fermion propagator
\begin{equation}
G^{^{\Lambda \xi }}(x-y)=\frac{1}{2\pi i}\frac{x\!\!\!\slash-y\!\!\!\slash}{%
(x-y)^{2}-i\varepsilon }\left( \frac{\tilde{\Lambda}_{\beta }^{2}}{M^{2}}%
\right) ^{\displaystyle\frac{1}{4}}\left[ \frac{{}}{{}}\!\!-M^{2}(x-y)^{2}+i%
\varepsilon \right] ^{\displaystyle\frac{1}{4}}  \label{th-eqx98}
\end{equation}%
where we have introduced an arbitrary mass scale $M$.

Immediately, we renormalize in configuration space by choosing the fermionic
wave function renormalization constant as being
\begin{equation}
Z_{\psi }^{^{\xi }}=\left( \frac{\tilde{\Lambda}_{\beta }^{2}}{M^{2}}\right)
^{\displaystyle\frac{1}{4}}\;.  \label{th-eqx99}
\end{equation}

Thus, in configuration space the renormalized propagator of massless
Thirring model in, its \textit{gauge invariant phase,} reads as
\begin{equation}
G^{^{R\xi }}(x-y)=\frac{1}{i2\pi }\frac{x\!\!\!\slash-y\!\!\!\slash}{%
(x-y)^{2}-i\epsilon }\left( \!\frac{{}}{{}}-M^{2}(x-y)^{2}+i\epsilon \right)
^{1/4}.  \label{th-eqx100}
\end{equation}%
Computing the Fourier transform we obtain renormalized fermion propagator as
being
\begin{equation}
\tilde{G}^{^{R\xi }}(p)=\frac{i}{p\!\!\!\slash}\left( \frac{\bar{M}^{2}}{%
p^{2}}\right) ^{\displaystyle\frac{1}{4}}{e}^{i\pi /4}\frac{\Gamma (5/4)}{%
\Gamma (3/4)}  \label{thb1-22}
\end{equation}%
This propagator deserves some comments. It does not have any
explicit dependence in the coupling constant nor in the ambiguity
parameter. Moreover, it is similar to the infrared behavior of the
fermionic propagator of the Schwinger model
\cite{schwinger,stam,radozycki1}, which has the confinement
phenomenon as one of their characteristics. And, the Schwinger model
presents chiral symmetry breaking due to the correlation functions
of the chiral densities break the cluster property which leads to
$\theta-$vacuum structure. Thus, \ two important quantum field
theoretical fenomena  could also be natural characteristics of the
massless Thirring model in its \textit{gauge invariant phase}.

%%%%%%%%%%%%%%%%%%%%%%%%%%%%%%%%%%%%%%%%%%%%%%%%%%%%%%%%%%%%%%%%%%%%%
%%%%%%%%%%%%%%%%%%%%%%%%%%%%%%%%%%%%%%%%%%%%%%%%%%%%%%%%%%%%%%%%%%%%%

\section{Remarks and conclusions}

%%%%%%%%%%%%%%%%%%%%%%%%%%%%%%%%%%%%%%%%%%%%%%%%%%%%%%%%%%%%%%%%%%%%%

The massless Thirring model was analyzed in a non--perturbative way
by using functional integral method. The key of the present analysis
is the calculus of the fermionic determinant which was made using
generalized prescription to control the UV divergences that appear
in its computation. This arbitrary procedure introduces an ambiguity
or Jackiw--Rajaraman parameter $a$ and it allows to obtain the new
solutions for the massless Thirring model.

\bigskip Next, we explain the contribution of the ambiguity for the new
solutions:\ the fermionic determinant is defined by the following functional
integral
\begin{equation}
\det \left( i\partial \!\!\!\slash+A\!\!\!\slash\right) =\int \!\!\mathcal{D}%
\psi \mathcal{D}\bar{\psi}\;\exp \left( i\int \!\!dx~\bar{\psi}\left(
i\partial \!\!\!\slash+A\!\!\!\slash\right) \psi \right)
\end{equation}%
if we make the following local phase transformation
\begin{equation}
\psi \rightarrow e^{i\rho \left( x\right) }\psi
%TCIMACRO{\TeXButton{space}{\quad,\quad}}%
%BeginExpansion
\quad,\quad%
%EndExpansion
\bar{\psi}\rightarrow \bar{\psi}{e}^{-i\rho \left( x\right) }
\end{equation}%
in our prescription the fermionic measure is not invariant under such
transformation and there is a nontrivial Jacobian given by \cite%
{Jackiwtop,Harada}
\begin{equation}
J_{\rho }\;=\;\frac{\det \left( i\partial \!\!\!\slash+A\!\!\!\slash\right)
}{\det \left( i\partial \!\!\!\slash+A\!\!\!\slash-\partial \!\!\!\slash\rho
\right) }\;=\;\exp \left( i\int \!\!dx\;\frac{a-1}{2\pi }\;A^{\mu }\partial
_{\mu }\rho -\frac{a-1}{4\pi }\;\partial _{\mu }\rho \partial ^{\mu }\rho
\right) .
\end{equation}%
It shows the explicit breaks of a local $U(1)$ symmetry which, in
general, is associated to the conservation of the fermionic current.
But such one local symmetry does not exist for the massless Thirring
model at classical level, thus, we are free to compute the fermionic
determinant in an arbitrary prescription characterized by the
ambiguity parameter $a$.

\bigskip At quantum level the Ward identity for the 1PI\ $2-$point for the
fermonic current, for $a\neq 1,$ allows to choose a value for the
ambiguity parameter in which it is possible to implement a local
$U(1)$ gauge symmetry for the model. This fact is reflected after
performing the fermionic integration, at this point it arises a
problem when we wish to perform the functional integration over the
auxiliary vector field for all values of the ambiguity parameter due
to there is a critical value $a=1-\displaystyle \frac{2\pi }{g}$
where it is not possible to be performed. For this value, the
effective action of the vector field gains a local $U(1)$ gauge
symmetry and we use the Faddeev--Popov technique to well--define the
functional integration. Consequently, we named as \textit{non--gauge
invariant phase} to the solutions with $a\neq 1-\displaystyle
\frac{2\pi}{g}$ and \textit{gauge invariant phase} to the solution
with $a=1-\displaystyle \frac{2\pi}{g}$. We also show that the
Green's functions, in both \textit{phases}, are solutions of the
Schwinger-Dyson equation for the massless Thirring model which
guarantees the validity of our approach.

We identify clearly the origin of the UV logarithmic divergences
present in the Green functions and make a non--perturbative
regularization into their generating functional via the
Pauli--Villars method. Such detailed analysis of the divergence
structure is fundamental to carry out the covariant
non--perturbative regularization of the the UV divergences, as it
was observed in \cite{trab1}. The UV divergences has a
non-perturbative character due to they can not be obtained via the
usual perturbative technique which uses the coupling constant as
parameter expansion; this divergence type has already observed in
the anomalous and chiral Schwinger model \cite{trab2}. Accordingly,
the regularization procedure modifies the components transversal and
longitudinal of the current--current interaction yielding us the
non--perturbative counter--terms which make possible the full
renormalization of the model in both phases.

A straightforward calculation for the higher fermionic correlation
functions from the regularized Lagrangian (\ref{th-eqx51}) and
(\ref{th-eqx94}) shows that they have already regularized. This fact
is notably important for the full renormalization of the model,
because it implies in the non existence of additional divergences in
the theory. Thus, in both phases, the coupling constant $g$ does not
renormalize such as it can be shown from the Ward's identities
(\ref{th-eqx34}) and (\ref{th-eqx87}) what was also observed in the
literature \cite{faber2}. The Green's functions for the composite
operators, as for example, chiral densities need an additional
renormalization procedure what will be reported in a following
paper.

In the \textit{non--gauge invariant phase} there are some values of
the ambiguity parameter that reproduce the solutions found in the
literature computed by different methods or techniques. On the other
hand, the \textit{gauge invariant phase} is a solution which is
implemented, for the first time, in the present work and it can not
be reproduced for no value of the ambiguity parameter or coupling
constant related to the non symmetric phase.

In the gauge invariant phase, we note that for a general value of
the gauge parameter $\xi$, the Green's functions are plagued of
infrared divergences but in the $\xi=0$ gauge they are infrared
finite, then, we can work in this gauge. On the other hand, the
non--symmetric phase is free of infrared divergences.

The gauge invariant phase shows some peculiar features, the first
feature is a simple equation (\ref{th-eqx12}) which relates the
ambiguity parameter $a$ and the coupling constant $g$. Possibly, it
could be fixed the ambiguity problem in the definition of the
fermionic current in the present model or given a guide to solve it
in the massive Thirring model \cite{fujita1}. This possibility comes
at quantum level where the model gets a local gauge symmetry which
allows to fix the ambiguity related to regularization freedom
originated from the classical level. The other one is the fermionic
propagator structure which resembles the behavior of the infrared
limit for the fermion propagator of the Schwinger model
\cite{schwinger,stam,radozycki1}. As it is well-known the Schwinger
model displays many characteristics of (1+3) dimensional more
realistic gauge theories, such as confinement or $\theta$--vacuum
structure, such characteristics could emerge in the present phase.
The study of the Hilbert space structure of the symmetric phase
would allow understanding if there is fermion confinement and would
identify the physical operators establishing clearly which
symmetries are present and which are violated at quantum level.
These questions mentioned above are at the moment under
investigation and results will be reported elsewhere.

%%%%%%%%%%%%%%%%%%%%%%%%%%%%%%%%%%%%%%%%%%%%%%%%%%%%%%%%%%%%%%%%%%%%%

\subsection*{Acknowledgements}

%%%%%%%%%%%%%%%%%%%%%%%%%%%%%%%%%%%%%%%%%%%%%%%%%%%%%%%%%%%%%%%%%%%%%

The author is grateful to Prof. C. R. Hagen for fruitful correspondence, and
Prof. R. Banerjee for comments on the present approach . I thank to Profs.
B. M. Pimentel and S. A. Dias for encouragement and fruitful comments. This
work was full supported by FAPESP/Brazil (grant 01/12611-7).

%%%%%%%%%%%%%%%%%%%%%%%%%%%%%%%%%%%%%%%%%%%%%%%%%%%%%%%%%%%%%%%%%%%%%

%%%%%%%%%%%%%%%%%%%%%%%%%%%%%%%%%%%%%%%%%%%%%%%%%%%%%%%%%%%%%%%%%%%%%


\begin{thebibliography}{99}
\bibitem{thirring1} W. Thirring, {Ann. Phys. (NY)} \textbf{3} (1958) 91

\bibitem{hagen1} C.R. Hagen, {Nuovo Cimento} \textbf{51B} (1967) 169.

\bibitem{klaiber1} B. Klaiber, in: A. Barut and W. Britten (Eds.), Lectures
in Theoretical Physics 1967, Gordon and Breach, NY, 1968, p. 141.

\bibitem{nakanishi} N. Nakanishi, {Prog. Theor. Phys.} \textbf{57} (1977)
580; 1025.

\bibitem{faber1} M. Faber and A. N. Ivanov, {Eur. Phys. J.} \textbf{C20}
(2001) 723.

\bibitem{faber2} M. Faber and A. N. Ivanov, {hep-th}/0112183.

\bibitem{faber3} M. Faber and A. N. Ivanov, {hep-th}/0206034.

\bibitem{Johnsonx} K. Jonhson, {Nuovo Cimento} \textbf{20} (1961) 773.

\bibitem{Scwingerx} J. Schwinger, {Phys. Rev. Lett.} \textbf{3} (1959) 296.

\bibitem{takahashi1} H. Takahashi and A. Ogura, {Prog. Theor. Phys.} \textbf{%
105} (2001) 495.

\bibitem{fujita1} T. Fujita, T. Kake, and H. Takahashi, {Ann. Phys.} \textbf{%
282} (2000) 100.

\bibitem{Jackiwtop} R. Jackiw, in: B.S. DeWitt and R. Stora (Eds.),
Topological Investigations of Quantized Gauge Theories, Relativity, Groups
and Topology II (Les Houches 1983), North Holland, Amsterdam, 1984.

\bibitem{banerjee1} R. Banerjee, {Z. Phys.} C25 (1984) 251; {Phys. Rev.}
\textbf{D37} (1988) 3778.

\bibitem{Harada} K. Harada and I. Tsutsui, Phys. Lett. \textbf{B183} (1987)
311.

K. Harada, H. Kubota and I. Tsutsui, Phys. Lett. \textbf{B173} (1987) 77.

\bibitem{DiasLinhares1} S. A. Dias e C. A. Linhares, {Phys. Rev.} \textbf{D45%
}, (1992) 2162.

\bibitem{mitra1} P. Mitra and A. Rahaman, {Ann. Phys.} \textbf{249} (1996)
34.

%%17

\bibitem{trab1} R. Casana and S. A. Dias, {Int. Jour. Mod. Phys.} \textbf{A15%
} (2000) 4603.

\bibitem{fujikawa1} K. Fujikawa, {Phys. Rev. Lett.} \textbf{42} (1979) 1195.

\bibitem{trab2} R. Casana and S. A. Dias, J. Phys. \textbf{G27} (2001) 1501.

\bibitem{Zinn-Justin} J. Zinn-Justin, {Quantum Field Theory and Critical
Phenomena}, $2^{\underline{\mbox{nd}}}$ edition, Oxford Science Pub., 1993.

\bibitem{Helayeldica} P. E. Haagensen and J. I. Latorre, Annals Phys.
\textbf{221} (1993) 77.

\bibitem{thirring2} W. Thirring, {Nuovo Cimento} \textbf{9} (1958) 1007 .

\bibitem{vglaser1} V. Glaser, {Nuovo Cimento} \textbf{9} (1958) 990.

\bibitem{shroer1} B. Schroer, {Fortschritte der Physik} \textbf{11} (1963) 1.

\bibitem{faddeev} L.D. Faddeev and V.N. Popov, {Phys.Lett.} \textbf{B25}
(1967) 29.

\bibitem{schwinger} J. Schwinger, {Phys. Rev.} \textbf{125} (1962) 397.

\bibitem{stam} K. Stam, J. Phys. \textbf{G9} (1983) L229.

\bibitem{radozycki1} T. Radozycki, {Eur. Phys. Journ.} \textbf{C6} (1999)
549.
\end{thebibliography}
\end{document}